%% file: main.tex
\documentclass[10pt,conference]{IEEEtran} 
\IEEEoverridecommandlockouts
\input{packages.tex}
\input{macros.tex}
\usepackage{cite}
\usepackage[utf8]{inputenc}
\usepackage[noend]{algpseudocode}
\usepackage{enumitem}
\def\BibTeX{{\rm B\kern-.05em{\sc i\kern-.025em b}\kern-.08em
    T\kern-.1667em\lower.7ex\hbox{E}\kern-.125emX}}
 
\title{Distribution of Quantum Circuits Over General Quantum Networks}

\author{\IEEEauthorblockN{Ranjani G. Sundaram}
\IEEEauthorblockA{\textit{Department of Computer Science} \\
\textit{ Stony Brook University}\\
 New York, USA}
\and
\IEEEauthorblockN{Himanshu {Gupta}}
\IEEEauthorblockA{\textit{Department of Computer Science} \\
\textit{ Stony Brook University}\\
 New York, USA}
\and
\IEEEauthorblockN{C. R. Ramakrishnan}
\IEEEauthorblockA{\textit{Department of Computer Science} \\
\textit{ Stony Brook University}\\
 New York, USA}

}

\begin{document}
\maketitle

\begin{abstract}
Near-term quantum computers can hold only a small number of qubits. One way to facilitate large-scale quantum computations is through a distributed network of quantum computers. In this work, we consider the problem of distributing quantum programs represented as quantum circuits across a quantum network of heterogeneous quantum computers, in a way that minimizes the overall communication cost required to execute the distributed circuit.  We consider two ways of communicating: cat-entanglement that creates linked copies of qubits across pairs of computers, and teleportation. The heterogeneous computers impose constraints on cat-entanglement and teleportation operations that can be chosen by an algorithm. We first focus on a special case that only allows cat-entanglements and not teleportations for communication. We provide a two-step heuristic for solving this specialized setting: (i) finding an assignment of qubits to computers using Tabu search, and (ii) using an iterative greedy algorithm designed for a constrained version of the set cover problem to determine cat-entanglement operations required to execute gates locally. 

For the general case, which allows both forms of communication, we propose two algorithms that subdivide the quantum circuit into several portions and apply the heuristic for the specialized setting on each portion. Teleportations are then used to stitch together the solutions for each portion. Finally, we simulate our algorithms on a wide range of randomly generated quantum networks and circuits, and study the properties of their results with respect to several varying parameters.


\end{abstract}

\input{Introduction}

\input{Background}

\input{ProblemFormulation}

\input{MemoryConstrained}
\input{DynamicPartitioning}

\input{Evaluation}

\section{Conclusion}
In this paper, we consider the problem of distributing a quantum circuit across a network of heterogeneous quantum computers in a way that minimizes the overall communication cost.  We described efficient algorithms which tackle issues that arise due to the heterogeneity of the network and different modes of communication.  We evaluated our algorithms on randomly-generated quantum circuits and networks to study their performance. Several avenues of future research remain.  Efficient simulation of quantum computations on classical machines (e.g.,~\cite{Wang-etal:Simulation:2021}), and analysis of quantum programs (e.g.,~\cite{Yu-Palsberg:2021}) employ partitioning similar to this work; it will be interesting to investigate this relationship further.  


\bibliography{main}
\end{document}

%% file: packages.tex
\usepackage{algorithm}
\usepackage{amsfonts}
\usepackage{amsmath}
\usepackage{color}
\usepackage{esvect}
\usepackage{gensymb}
\usepackage{graphicx}
\usepackage{mathtools}
\usepackage[binary-units=true]{siunitx}
\usepackage{stfloats}
\usepackage{url}
\usepackage{xspace}
\usepackage{wrapfig}
\usepackage[normalem]{ulem}
\usepackage{subcaption}

%% file: macros.tex
\newcommand{\eat}[1]{}

\newcommand{\blue}[1]{{\color{black}#1}}

\newcommand{\para}[1]{\smallskip \noindent {\bf #1}}
\newcommand{\softpara}[1]{\smallskip \noindent \underline{#1}}

\newcommand{\showOutline}{s}

\definecolor{darkgreen}{rgb}{0,0.5,0}
\definecolor{darkblue}{rgb}{0,0,0.5}

\newcommand{\outline}[2]{\if#1\showOutline {\color{darkgreen}#2} \fi}
\newcommand{\detailedoutline}[2]{\if#1\showOutline {\color{darkblue}#2} \fi}

\newcommand{\todo}[2]{\if#1\showOutline {#2} \fi}

\newcommand{\purp}[1]{} 

\def\blackbox{\hfill {\vrule height6pt width6pt depth0pt}}
\newcommand{\vsplem}{\vspace{0.025in}}
\newcommand{\vsp}{\vspace*{-0.1in}} 

\newtheorem{lem}{Lemma}
\newtheorem{thm}{Theorem}
\newtheorem{obv}{Observation}

\newenvironment{lem-wo-prf}{\begin{lem} \nopagebreak}{{\hfill$\blackbox$} \end{lem}}
\newenvironment{lem-w-prf}{\vsp \begin{lem} \nopagebreak}{\end{lem}}
\newenvironment{thm-w-prf}{\vsp \begin{thm} \nopagebreak}{\end{thm}}
\newenvironment{thm-wo-prf}{\begin{thm} \nopagebreak}{{\hfill$\blackbox$} \end{thm}}

\newcounter{packednmbr}

\newcommand{\aza}{{\tt AG-Algo}\xspace}
\newcommand{\tabu}{{\tt Tabu}\xspace}

\newcommand{\eps}{\mbox{EP}\xspace}
\newcommand{\epss}{\mbox{EPs}\xspace}

\newcommand{\id}[1]{\textit{#1}}
\newcommand{\cz}{\id{CZ}\xspace} 

\newcommand{\dqc}{{DQC}\xspace}
\newcommand{\dqcm}{{DQC-M}\xspace}

\newcommand{\gdqc}{{DQC}\xspace}

\newcommand{\nta}{{\tt DQC-M}\xspace}
\newcommand{\ntag}{{\tt DQC-M-Greedy}\xspace}
\newcommand{\seq}{{\tt Sequence}\xspace}
\newcommand{\slt}{{\tt Split}\xspace}

\newcommand{\ts}{\ensuremath{t_s}\xspace} 
\newcommand{\te}{\ensuremath{t_e}\xspace} 
 
\newcommand{\qi}{\ensuremath{q_i}\xspace} 
\newcommand{\qj}{\ensuremath{q_j}\xspace}

\newtheorem{defin}{Definition}
\newenvironment{definition}{\begin{defin}} {{\hfill$\Box$}\end{defin}}

%% file: Introduction.tex
\section{\bf Introduction}


\para{Motivation.} There are two crucial technological hurdles that constrain the realization of quantum computing's  potential:  (a) the limited number of \emph{qubits}, the basic store of quantum information, in any single quantum computer; (b) severe loss of information due to noisy operations and unwanted interactions with the environment~\cite{Preskill2018quantumcomputingin}.  The second hurdle can be overcome in principle by using error-correcting codes, but that results in a blowup in the number of qubits needed for a computation, thereby exacerbating the first hurdle.  
Distributing a quantum computation requiring a large number of qubits over a network of quantum computers (QCs) is a way to overcome this \blue{hurdle}~\cite{cirac1999distributed, YimsiriwattanaL:04,YimsiriwattanaL:05}.

\para{State of the Art.}
Quantum \emph{circuits}, which specify a sequence of \emph{gates} (operations) on a set of qubits, is a common abstraction between higher-level quantum \emph{programs} and lower-level computing hardware.  Distributing a quantum circuit over a quantum network involves assigning the circuit's qubits to QCs, and introducing communication operations to perform non-local operations (i.e. operations that span multiple QCs).  The cost of distributing a circuit is the number of added communication operations.

Optimally distributing a given quantum circuit for evaluation over a network of QCs has been the focus of several earlier works~\cite{Daei+:19,Andres-MartinezH:19,g2021efficient}.  They assume a \emph{homogeneous} network--- all QCs in the network have the same number of qubits, and the cost of quantum communication between any pair of QCs in the network is uniform.  Even under this setting, the optimal distribution problem is intractable~\cite{Andres-MartinezH:19}.

While~\cite{Daei+:19} use teleportation as the only means of communication between QCs, \cite{Andres-MartinezH:19} and~\cite{g2021efficient} use \emph{cat-entanglement}~\cite{YimsiriwattanaL:04} which allows for the creation of shared copies of qubits (see \S\ref{sec:bg-qc}) and often yields lower-cost solutions. But~\cite{Andres-MartinezH:19} and~\cite{g2021efficient} ignore the storage requirements for multiple simultaneous cat-entanglements which could be substantial~\cite{Andres-MartinezH:19}.  \blue{Distributed quantum computation over heterogeneous networks is considered in~\cite{Ferrari-etal-2021}, where the cost, measured as increase in circuit depth, is not minimized, but bounded by a linear factor.} 

\para{This Paper.}   
In this work, we consider the problem of \emph{optimally distributing a quantum circuit across an \blue{arbitrary} topology network of heterogeneous quantum computers.}  In particular, we consider the following generalizations to the optimal distribution problem:
\begin{enumerate}
    \item The cost of communication between two QCs in a given network is a function of their network distance. 
    \item Each QC has specified qubit capacity, and an ``execution memory" of limited size for storing cat-entangled qubits; these two limits may vary across QCs in the network.
    \item Communication may be via cat-entanglement, or teleportation whose effect is to dynamically alter the assignment of qubits to QCs. 
\end{enumerate}
Performing distributed quantum computing in practice, when technology makes it feasible, will require us to drop the homogeneity assumption and study the problem in the more general setting described above. 

\para{Contributions and Organization.}
We formalize the optimal distribution problem under these generalizations as the \gdqc problem.  We then provide polynomial-time heuristics for this intractable problem in multiple steps. 

We first consider a special case of \gdqc, called \dqcm, that considers only cat-entanglement-based communication (i.e., without generalization (3) above); see \S\ref{sec:mdqcproblem}.  We solve \dqcm in two steps: (i) First, we use a Tabu-search-based heuristic to partition the given circuit's qubits among QCs taking the heterogeneity of the network and storage limits into account.
(ii) Then, we develop an algorithm for introducing cat-entanglements to ``cover''  non-local gates, i.e., gates whose operands are in assigned to QCs.  In a restricted setting, this yields an $\mathcal{O}(\log n)$-approximate solution (here $n$ is the number of non-local gates). 

For solving the general \dqc problem (see \S\ref{sec:dynamicPart}), we provide two greedy heuristics, \seq and \slt, each using our solution to the \dqcm problem as a subroutine.   
Although \seq has lower complexity than \slt, neither is uniformly better than the other---we provide examples where each heuristic outperforms the other. In \S\ref{sec:eval}, we present our evaluation results, which show that \slt performs better than \seq in most cases.

We begin with a brief background in quantum computation and communication (\S\ref{sec:bg-qc}) for completeness followed by a formal description of the problem (\S\ref{section:dqcproblem}).

%% file: Background.tex
\section{\bf Background: Quantum Computation and Communication}
\label{sec:bg-qc}

We start with giving a brief background on two key quantum concept relevant to our paper: quantum circuits and our choice of universal gate set as unary and CZ gates, and quantum 
communication methods used in our work.

\para{Quantum Circuits.}  Quantum computation is typically abstracted as a \emph{circuit}, where horizontal ``wires'' represent \emph{qubits} which carry quantum data, and operations on the qubits performed by vertical ``gates'' connecting the operand wires~\cite{nielsen_chuang_2010}.  Quantum computers (QCs) evaluate a circuit by applying the gates in the left-to-right order, so this circuit can also be understood as a sequence of machine-level instructions (gates) over fixed number of data cells (qubits).   

\begin{figure}[t]
\centering
\includegraphics[width=0.5\textwidth]{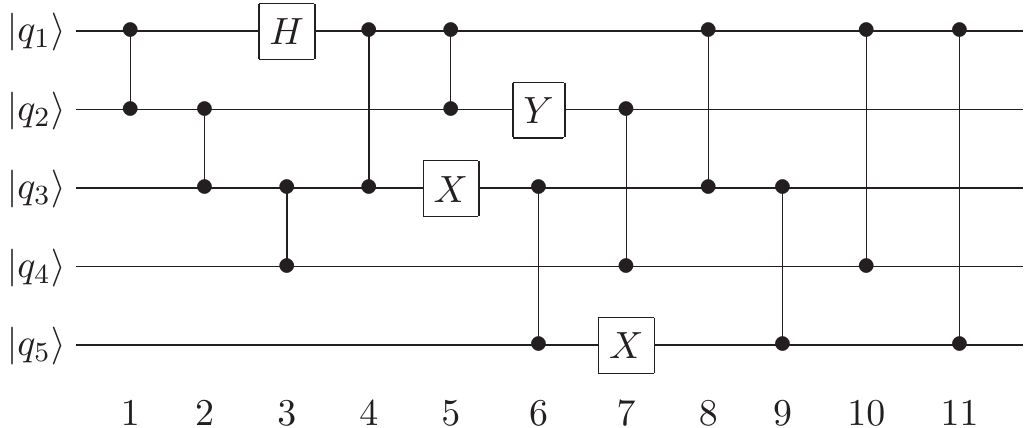}
\caption{Quantum Circuit Example.}
\vspace*{-3ex}
\label{fig:bg-circuit-eg}
\end{figure}

Analogous to classical Boolean circuits, there are several universal gate sets for quantum computation: any quantum computation can be expressed by a circuit consisting only of gates from a universal gate set. In particular, the ``\emph{Controlled-Z}'' binary gate, denoted by \cz, along with the set of all possible unary gates forms a universal gate set. We use this universal gate set in this paper since the symmetry of \cz gates allows a simpler formulation of   distributed execution by creating linked copies using cat-entanglements (see below).
Fig.~\ref{fig:bg-circuit-eg} shows the pictorial representation of an example circuit, consisting only of unary gates (boxes) and \id{CZ} gates (vertical connectors). 
Without loss of generality, we ignore measurement gates; measurement can be postponed to the end and treated as unary operations. 

\para{Quantum Communication.}
If a given quantum circuit is to be evaluated in a distributed fashion over a network of QCs, we have to first distribute the qubits over the QCs.  But such a distribution may induce gates in the circuit to span different QCs.  To execute such \emph{non-local} gates, we need to bring all operands' values into a single QC via quantum communication. 
However, direct/physical transmission of quantum data 
is subject to unrecoverable errors, 
as classical procedures such as amplified signals 
or re-transmission cannot be applied due to quantum no-cloning~\cite{wooterszurek-nocloning,Dieks-nocloning}.\footnote{Quantum error
correction mechanisms~\cite{muralidharan2016optimal,devitt2013quantum} can be used to mitigate the transmission errors, 
but their implementation is very challenging and is not expected to be used
until later generations of quantum networks.}
Fortunately, there are other viable ways to 
communicate qubits across network nodes, as described below.



\softpara{Teleportation.}
An alternative approach to physically transmit qubits is via \emph{teleportation}~\cite{Bennett+:93} 
which requires an a priori distribution of maximally-entangled pair (MEP) of qubits (e.g., Bell Pair) over the two nodes.  With an MEP distributed over nodes $A$ and $B$,  teleportation of a qubit state from $A$ to $B$ can be accomplished using classical communication and local gate operations, while consuming/destroying the MEP. 

\softpara{Cat-Entanglement: Creating ``Linked Copies'' of a Qubit.}
Another means of communicating qubit states is by creating \emph{linked copies} of a qubit across QCs, via \emph{cat-entanglement} operations~\cite{Eisert+:00,YimsiriwattanaL:05} which, like teleportation, require
a Bell Pair to be shared \emph{a priori}. These linked copies are particularly useful in efficient distributed evaluation of circuits involving only \cz and unary gates, as follows. 
The symmetry of \cz operation allows for either of the qubit operands to act as the (read only) control operand, and, more importantly, the control qubit can be just a linked copy of the original qubit operand (and since a linked copy is read-only, many  copies can exist and used simultaneously). 
However, since a unary operation on the original qubit $q$ may \emph{change} its state, 
linked copies of $q$ may not remain true copies; thus, we ``disentangle'' any linked copies of $q$ via a dual operation called \emph{cat-disentanglement} before applying a unary operation on $q$---the dis-entanglement operation doesn't require a Bell Pair.


%% file: ProblemFormulation.tex
\section{\bf Relevant Concepts and Problem Formulation}
\label{section:dqcproblem}

In this section, we define the \dqc problem of distributing quantum circuits across 
quantum computers. We start with an informal description, define the relevant terms
and concepts, and then formulated the \dqc problem addressed in this paper. 

\para{Informal Problem Description.}
The goal of the {\em Distributing Quantum Circuits} (\dqc) problem addressed in this paper
is to determine an efficient distribution of a given quantum circuit, over a given 
network of QCs. Efficient distribution essentially entails two tasks: distributing the qubits
over the distributed QCs, \blue{and then executing the given gates, including non-local gates using a judicious combination of teleportation and/or cat-entanglement operations.}
Informally, the \dqc problem is to execute the given (centralized) quantum circuit over the
given quantum network using a minimum cost of teleportations and cat-entanglements used
to execute the non-local gates, under the given memory constraints. 


\para{Closest Related Work.} The closest work that addresses the above problem is our own recent work~\cite{g2021efficient} ---where we address the \dqc problem under the simple settings of homogeneous computers with unbounded execution memory (to store cat-entanglement copies), complete network topology, and no teleportations. 
For the simplified setting,~\cite{g2021efficient} presents a two-step algorithm for the \dqc problem, wherein
the first step determines the partitioning of qubits to computers through balanced graph partitioning and the second step minimizes the number of cat-entanglement operations via
an iterative greedy approach. 

In this paper, we address the generalized \dqc problem wherein each computer may have non-uniform 
storage memory (to store the qubits) and bounded non-uniform execution memory (to allow for copies
from cat-entanglements). Most importantly, we allow teleportations, which may dynamically change the partitioning of qubits across computers, but can improve the communication cost.

\subsection{\bf Key Concepts and Terminology}
\label{sec:terms}

\para{Quantum Circuit Representation.} 
As in~\cite{Andres-MartinezH:19}, we consider the universal gate set with (binary) \cz and unary gates. 
Also, in our context, we do not need to represent the type of unary gates.  Thus, we represent an \emph{abstract quantum circuit} $C$ over  a set of qubits $Q = \{q_1, q_2, \ldots\}$ as a sequence of gates $\langle g_1, g_2, \ldots \rangle$ where each $g_k$ is either binary CZ gate
or a unary gate. 
We thus represent binary gates in a circuit as triplets $(q_i, q_j, k)$, where $q_i$ and $q_j$ are the two operands, and $k$ is the time instant (see below) of the gate in the circuit; and unary gates as pairs $(q_i, k)$, where $q_i$ is the operand and $k$ is the time instant.
We use $N_q$ and $N_g$ to denote the number of qubits and gates in the circuit, respectively.

\begin{figure}[t]
\centering
\includegraphics[width=0.48\textwidth]{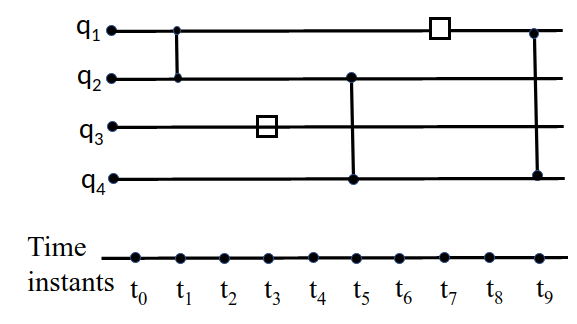}
\caption{In the above figure, \blue{the gates are at odd-numbered time instants} $t_1, t_3, \ldots, t_{9}$, and the even-numbered time instants between each pair of consecutive gates have been introduced for convenience.}

\vspace*{-3ex}
\label{fig:time_instants}
\end{figure}

Each gate occurs uniquely at a time instant.
In addition to the instants where the gates occur, we introduce additional {\bf time instants} in between gates for cat-entanglement/teleportation operations. See Fig.~\ref{fig:time_instants}.

\para{Quantum  Network (QN).}
We represent quantum network as a connected undirected graph with nodes representing QCs and edges representing  (quantum and classical) direct communication links. \blue{Nodes of the network are denoted by $P$;} we use the words node, computer, and QC interchangeably.  \blue{We denote the number of nodes in the network by $N_p$.}
Each computer $p\in P$ has quantum memory to store qubits; for simplicity, we divide this memory into two parts: qubit storage
memory to store the ``original'' qubits, with capacity denoted by $s_p$, and \emph{execution memory} used to store the linked copies (ebits) from cat-entanglements, with capacity denoted by $e_p$. Thus, as part of the given QN specification, 
each node has 
has a certain amount of qubit storage and execution memory. 

\para{Home Computers.}
To distribute execution of a given quantum circuit, we first distribute qubits of the given circuit across the network nodes.
At any point of time, each qubit $q$ of the circuit resides (i.e., is stored) at a unique node in the network---which we call its \emph{home computer} or just \emph{home}.
Cat-entanglement will create a linked copy of a qubit $q$ at another computer, but does not change $q$'s home. However, a qubit's home can be changed by teleporting it to another computer. 

The home computers of qubits are represented by a set of home-computer functions, $\pi_t$, one for each time
instant $t$.  Each home-computer function
maps a circuit's qubits to computers, i.e., $\pi_t: Q \mapsto P$. Thus, $\pi_t(q)$ denotes the home of qubit $q$ at time $t$. A home-computer function $\pi_t$ is valid if and only if it obeys the storage memory constraint--- i.e., for any computer $p$ with a storage memory of $s_p$ units,
there are at most $s_p$ qubits $q$ such that $\pi_t(q) = p$. A gate $(q_i, q_j, t)$ is defined as \textbf{non-local} at time $t$ if $q_i$ and $q_j$ have different home-computers, i.e., $\pi_{t}(q_i) \neq \pi_{t}(q_j)$.

\para{Representing Teleportations.}
We represent teleportation by a triplet $(q, p, t)$ which signifies that the qubit $q$ was teleported to 
the computer $p$ at time $t$. The teleportation $(q, p, t)$ results in changing
the home-computer function such that $\pi_t(q) = p$. For simplicity, we enforce that teleportations only happen at times with no gates.
To simplify the issue of storage memory violations due to teleportations, we assume that all the
teleportation at time $t$ happen simultaneously- -- and do not require additional
memories for the \epss used in teleportations. Thus, the set of teleportations occurring at time $t$ can be 
looked upon as changing the entire home-computer function from a valid $\pi_{t-1}$ to a valid $\pi_{t}$. 

\para{Migrations (formalizing Cat-Entanglements).}
As described before, we use
cat-entanglements to make linked copies of qubits to execute non-local gates.
As in our earlier work~\cite{g2021efficient}, we use the term {\em migrations} to denote cat-entanglement.
However, since we now allow teleportations
which change the home computers of qubits over time, formal definition of
migrations differs from that used in~\cite{g2021efficient}. 
Informally, a qubit $q$ can be migrated from its 
home computer to another for a certain duration of time $(\ts, \te)$; such a migration is considered
 {\em valid} if, during the time interval $(\ts, \te)$,  there are no unary operation on $q$ and its home computer doesn't change. For simplicity, we assume that there are no gates at $\ts$ and $\te$. 

\begin{definition}[Migration]
\label{defn:migration}
Given a quantum circuit, a quantum network, and the home-computer function at each time instant, 
a \emph{migration} is a quadruple $(q, p, t_s, t_e)$ to denote migration of qubit $q$ from its home-computer at $t_s$ to another computer $p$ for the period $(t_s, t_e)$. For the migration to be valid,
the following conditions must hold. 
\begin{itemize}
    \item $p \not=\pi_{t_s}(q_i)$, i.e., $q$ is migrated to a computer $p$ different from its home computer at $t_s$.
    \item $\pi_t(q)=\pi_{t_s}(q)$ for all $t$ in $(t_s, t_e)$. That is, the qubit $q$'s home computer doesn't change for the duration of the migration; i.e., $q$ is not teleported during the period.
    \item There are no unary gates on $q$ during  $(t_s, t_e)$.
\end{itemize}  
\vspace*{-15pt}
\end{definition}

\softpara{Coverage by a Migration; Home-Coverage.} \label{defn:coverage}
We use the term ``cover'' to denote  migrations that help execute a non-local gate, and formally define the notion of coverage of 
a gate by migration(s) as follows. 
A binary gate  $g=(\qi, \qj, t)$ can be covered by one or two migrations as follows.
\begin{enumerate}
    \item By a single migration $(\qi, \pi_t(\qj), t_s, t_e )$ or $(\qj, \pi_t(\qi), t_s, t_e)$, where $t_s\leq t \leq t_e$; this represents migrating one operand to the other's home computer to enable the gate's local execution.
    \item By a pair of migrations \{$(\qi, p, t_{s1}, t_{e1}), (\qj, p, t_{s2}, t_{e2})\}$ for some computer $p$, where $t_{s1} \leq t \leq t_{e1}$ and $t_{s2} \leq t \leq t_{e2}$. This represents migrating both operands to a common computer $p$ and executing the gate locally there.
\end{enumerate}
Coverage of a gate by a single migration 
is called \emph{home-coverage}.   

\para{Feasible Set of Migrations.}
Limited execution memory at each computer restricts the maximum number of linked copies that can be present in a computer at any point of time. 
Consequently, a set of migrations is
feasible only if the created linked copies obey 
the execution memory constraint at every computer at every point of time. We define this formally below.

Let $m$ be a migration, $p$ a computer, and $t$ a time instant, and let $\mathcal{A}(m, p, t)$ be a function that is $1$ iff there is a linked-copy of a qubit at computer $p$ at time $t$ due to migration $m$. More formally:%
\begin{align*}
    \mathcal{A}(m, p,  t)=
    \begin{cases}
    1 \text{ \quad if $m = (q, p, t_s, t_e)$ and $t_s\leq t \leq t_e$}\\
    0 \text{ \quad Otherwise}
    \end{cases}
\end{align*}

A set of migrations $\mathcal{M}$ are said to be \emph{feasible} if and only if $\sum _{m \in \mathcal{M}} \mathcal{A}(m, p, t) \leq e_p$ for all computers $p$, for all times $t$, where $e_p$ is the capacity of execution memory at $p$.

\para{Cost of Migrations and Teleportations.} 
The cost of a migration $(q, p, t_s, t_e)$ is 
defined as the distance between the nodes $\pi_{t_s}(q)$ to $p$ in the given network graph.
This cost accounts for the fact that migrating a qubit from $\pi_{t_s}(q)$ to $p$ requires
an \eps over nodes $p$ and $\pi_{t_s}(q)$ whose generation cost we assume to be proportional
to the distance between $p$ and $\pi_{t_s}(q)$. 
Similarly, cost of a teleportation $(q,p,t)$ is defined as to be the distance between $\pi_{t-1}(q)$ to $p$.

\subsection{\bf Problem Formulation and Example}

We now define the \gdqc problem formally, based on the above concepts and terms.

\para{\gdqc Problem.}
Given a quantum circuit and a quantum network, the \gdqc problem is to:
(i) determine a valid home-computer function at all time instants
(which also yields the teleportations incurred),
and (ii) a feasible set of migrations that cover all the non-local gates,
while minimizing the total cost of migrations and teleportations used.

The above \dqc problem can be shown to be NP-hard, by a reduction from the \dqc problem
that only allows migration (and no teleportations) which is known to be NP-hard~~\cite{Andres-MartinezH:19}. We omit the details of the 
reduction here.

\begin{figure}
\centering
\includegraphics[width=0.48\textwidth]{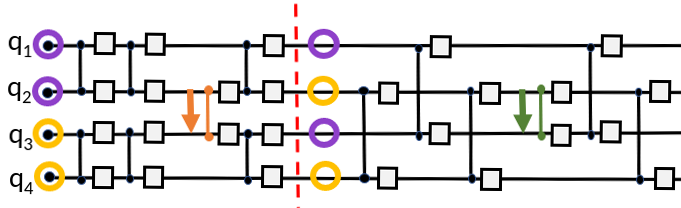}
\eat{
\caption{\blue{A} \dqc problem instance with four qubits and two computers. 
The figure also shows \blue{an optimal} solution, 
with the colored circles signifying the home-computer function, and the two arrows signifying migrations, \blue{and the dashed vertical line signifying teleportations that interchange the home computers of qubits $q_2$ and $q_3$.}}}
\caption{\dqc problem instance from Example~1}
\vspace*{-3ex}
\label{fig:runningExample}
\end{figure}

\para{Example 1.} Consider a \dqc problem instance in Figure~\ref{fig:runningExample}---a circuit with four qubits and two computers each with a storage memory of two and execution-memory of 1. We assume the computers to be connected by a network link, and thus, the cost of any migration or teleportation is one.
The figure also illustrates an optimal solution to the \dqc problem of cost four. Initially, qubits $q_1$ and $q_2$ are assigned to the first computer (signified by purple circles), while $q_3$ and $q_4$ are assigned to the second computer (signified by yellow circles). At the time instant denoted by the red line, the qubit assignment is changed (via appropriate teleportations): $q_2$ is teleported to the second computer while $q_3$ is teleported to the first. The only non-local gates are the ones marked in orange and green --- each requiring one migration. Thus, the total cost is 4, comprised of two teleportations and two migrations.
 
    
    

%% file: MemoryConstrained.tex
\section{\bf \dqcm Problem: \dqc with Only Migrations}
\label{sec:mdqcproblem}

For ease of presentation, we first address a simpler version of the \gdqc problem, wherein we do
not allow any teleportations. We refer to this problem as \dqcm. In this special case \dqcm problem, the home-computers of the qubits never change after the initial placement and we need to cover all the non-local gates with just migrations. 
In effect, the \dqcm problem boils down to picking an initial assignment of qubits to computers such that non-local gates can be covered with minimum-cost migrations. 
We design a two-step algorithm for the \dqcm problem, as discussed below. 

\para{Two-Step Algorithm (\nta) for \dqcm.}
The \dqcm  problem is a direct generalization of the problem addressed in our earlier
work in~\cite{g2021efficient} wherein the computers were assumed to have uniform storage memory with unbounded execution memory and the network was assumed to have a complete topology.  
As in~\cite{g2021efficient}, we develop a two-step algorithm, we call \nta,
wherein in the first step we determine the assignment of qubits to computers (i.e., the
initial home-computer mapping which then remains unchanged),  and then, in the second step,
we determine
the migrations to cover the non-local gates. We discuss these two steps in the following subsections.
See the high-level pseudo-code of the two-step \nta algorithm.

\floatname{algorithm}{Algorithm}
\begin{algorithm}[t]
\caption{\nta}
  \label{alg:NoTele}  
  \begin{flushleft}
  \textbf{Input:}  Quantum circuit $C$ over qubits $Q$, Network graph $G$ with nodes $P$\\
  \textbf{Output:} A valid home-computer function $\pi$ and a set of feasible migrations $\mathcal{M}^*$ that cover all non-local gates in $C$
  \end{flushleft}
  \begin{algorithmic}[1]
        \State $\pi \leftarrow$ A valid home-computer function such that the cost required migrations is low.
        \State ${\mathcal{M}}\leftarrow$ A low-cost set of migrations that covers all the non-local gates resulting from $\pi$.
        \State (Post-processing step) $\mathcal{M}^* \leftarrow$ A low-cost set of feasible migrations covering all non-local gates obtained by resolving memory constraint violations in $\mathcal{M}$.
        \State \Return $\pi, {\mathcal{M}^*}$ 
  \end{algorithmic}
\end{algorithm}

\input{NetworkTop}

\subsection{\bf \nta Step 2: Selection of Migrations to Cover Gates}
\label{sec:step2-wo-tel}

In this section, given an assignment of qubits to computers, we seek to compute a minimum-cost feasible set of migrations that cover all the non-local gates.

\para{Basic Idea.}
\blue{Selecting migrations to cover non-local gates is essentially 
a generalization of the set-cover problem,  with two key differences:} 
\begin{enumerate}
\item
First, we are restricted to choose only a feasible set of migrations. 
Fortunately, the execution-memory constraints can be expressed as linear constraints---and hence, can be handled by using approximation techniques from~\cite{azar2012efficient} that studies the related maximum-coverage problem with linear constraints.
\item
Second, a gate may be covered by a pair of migrations together which translates to allowing a pair of sets to cover an element together; see the ``{\em Coverage by a Migration}'' paragraph in \S\ref{sec:terms}.
Such a generalization breaks the submodularity of the objective function---and in general, can render the coverage problem inapproximable. 
\end{enumerate}
\blue{It should be noted that with generalization (2) alone but in the absence of (1), we gave an approximation algorithm in~\cite{g2021efficient}.}
However, neither the technique from~\cite{azar2012efficient} nor~\cite{g2021efficient} can be extended to 
handle both the above generalizations together while ensuring a performance guarantee. 
Thus, we develop a heuristic based on ~\cite{azar2012efficient} as described below. We start with considering the special case of home-coverage (i.e., select migrations to cover gates using only home-coverage), and then extend our algorithm to general coverage.

\floatname{algorithm}{Algorithm}
\begin{algorithm}[t]
\caption{Step 2 of \nta.}
  \label{alg:step-2-dqcm}  
  \begin{flushleft}
  \textbf{Input:} Quantum Network $G$, Quantum circuit $C$,  Home-computer function $\pi$. \\
  \textbf{Output:} A set of migrations $\mathcal{M}$ that covers all non-local binary gates.
  \end{flushleft}
  \begin{algorithmic}[1]
        \State uncovered $\leftarrow$ non-local binary gates in ${C}$ due to $\pi$.
        \While{(uncovered) }
            \State ($S$, covered) $\leftarrow$ \textsc{Cover-$\alpha$}(uncovered, $G$, $\pi$)
            \State uncovered $\leftarrow$ uncovered $\setminus$ covered
            \State $\mathcal{M} \leftarrow \mathcal{M} \cup S$
        \EndWhile 
        \State \Return $\mathcal{M}$.  
        \Function{Cover-$\alpha$}{uncovered, $G$, $\pi$}
         \State minCost = 1; 
         \State maxCost = $|$uncovered$|$ $\times$ (Diameter of $G$)
         \For {c in [minCost, maxCost]} 
         \State ($S$, covered) $\leftarrow$ \aza(c, ${C}$, $\pi$, $G$, uncovered)  
         \If{$|$covered$|\geq \alpha|$uncovered$|$}
         \State Break
         \EndIf
         \EndFor
         \State \Return ($S$, covered)
        \EndFunction
  \end{algorithmic}
\end{algorithm}

\para{Approximation Algorithm for Home-Coverage.} 
We note that \textsc{Multiplicative-updates} algorithm from~\cite{azar2012efficient}, hereafter referred to as the \aza, 
maximizes the number of elements covered under a {\em given cost
budget} of sets with linear constraints. In contrast, the Step-2 of \nta needs to select minimum-cost migrations to cover
{\em all} gates---which is in some sense, a dual of the \eat{maximum-coverage} problem solved by \aza. 
To cover all the gates with minimum-cost migrations based on \aza, we use an iterative algorithm where, in each iteration, we cover at least a certain constant fraction $\alpha$ of the remaining gates using a minimum-cost set of migrations.  Iterations are repeated until all binary gates are covered.  \blue{To find a minimum-cost set of migrations covering at least $\alpha$ fraction of the  gates using \aza, we exploit the fact that cost is an integer bounded by the product of the number of binary gates in a circuit and the diameter of the network, as shown below.}
\eat{
At a high-level, our migration-selection algorithm for the second step of \nta is as follows.
\begin{itemize}
    \item In each iteration, pick a minimum-cost set of feasible migrations that cover at least $\alpha$ fraction of the remaining uncovered binary gates, as described below.
    \item Repeat until there are no uncovered binary gates left.
\end{itemize}
Each iteration to cover an $\alpha$ fraction of remaining gates works as follows.
}
\begin{itemize}
    \item For each cost $c$:
    \begin{itemize}
        \item Select, using \aza, a feasible set of migrations costing at most $c$ that covers the maximum number of gates.
    \end{itemize}
    \item
    Pick the solution with smallest $c$ for which the \aza solution could cover $\alpha$ fraction of the remaining gates. 
\end{itemize}

We use $\alpha = 0.4$ in our implementation, based on the approximation factor of the \aza.
For a more formal and complete description, see the pseudocode shown in Algorithm~\ref{alg:step-2-dqcm}. We make two remarks. First, \blue{while each iteration returns a feasible set of migrations, their union may not be feasible,} i.e., may violate execution-memory constraints; we resolve them as a post-processing step below. Second,
in subroutine \textsc{Cover-$\alpha$}, we can use a binary search to more efficiently iterative over all possible costs.

For sake to clarity, we have intentionally omitted details of the \aza algorithm, but at a high-level it is an iterative approach that picks the migration based on an objective that considers both---the number of gates covered as well as the ability to cause constraint violations.

\softpara{Performance Guarantee.} It can be shown that the above algorithm yields a $\mathcal{O}(\log n)$-approximation solution (where $n$ is the number of non-local gates) for the problem of selection of minimum-cost set of migrations to cover all gates given a home-computer function,
\blue{ while bounding the violation of execution-memory constraints (violations fixed in \S\ref{sec:post-proc})}.  Note that $n \ll N_g$, the number of all gates in the circuit.
We formalize the performance guarantee below. 

\begin{thm-wo-prf} 
Given a network $G$, circuit ${C}$, and a home-computer function $\pi$, let $k^*$ be the optimal-cost of a set of feasible migrations that home-covers all the non-local gates for $\pi$, and $n$ be the total number of non-local gates for $\pi$. 
\blue{Algorithm~2} returns a solution $\mathcal{M}$ such that: 
\begin{itemize}
    \item $\mathcal{M}$ covers all non-local gates.
    \item $|\mathcal{M}|\leq (\log n) k^*$. 
    \item \blue{For every computer $p$ in $G$, the amount of execution memory in $p$ used  by ${\cal M}$ at any time instant is at most $(\log n) e_{p}$.} 
\end{itemize}
\vspace*{-15pt}
 \end{thm-wo-prf}
 
 \para{Generalization to General Coverage.}
Recall that the above algorithm was under the restriction of home-coverage. To allow for general coverage of gates by migrations, i.e., to allow a pair of migrations to together cover a gate,  we need to modify the \aza subroutine accordingly. Note that \aza works iteratively, wherein in each iteration it selects a single migration. To allow for general coverage, we modify the \aza subroutine to also consider pairs of 
migrations for selection in each iteration. \blue{This change is straightforward,} and we omit the details. Unfortunately, allowing general coverage by migrations breaks down the approximation guarantee of \aza.

\subsection{\bf Post-Processing to Resolve any Memory Violations.}
\label{sec:post-proc}

While our algorithm selects a feasible set of migrations in every iteration, the overall solution may violate memory constraints. We resolve these violations by replacing some migrations with migrations of smaller duration.
Consider a migration $(q, p, t_s, t_e)$ which covers gates at time $t_1, t_2$ and $t_3$. We can covert this migration into three separate migrations, viz., $(q, p, t_1, t_1), (q, p, t_2, t_2)$ and $(q, p, t_3, t_3)$, each of which covers the gates at $t_1, t_2$ and $t_3$ respectively. 
The above conversion reduces the usage of execution memory at $p$, while increasing the total cost of migrations. 
Note that there always exists a solution that uses only such ``instanteneous'' migrations and requires only one unit of execution memory at each computer, since there is at most one gate at each time instant. 
Thus, a simple strategy to resolve execution-memory violations could be to
convert migrations into multiple shorter migrations iteratively.\footnote{Note that since our algorithms only create migrations for yet-uncovered gates, such a conversion strategy would not yield ``redundant'' instantaneous migrations---and thus, conversions alone should yield a feasible solution.}

Based on the above, our post-processing algorithm to resolve execution-memory violations is
as follows: we iteratively pick the migration that causes a violation and covers the least number of gates, and covert it into instantaneous migrations as described above. 

\para{Time Complexity of \nta Algorithm.} Overall, the \nta Algorithm runs in
$\mathcal{O}(\lambda {N_q}^3 + N_pn^6\log n)$ time, where $\lambda$ is the number of iterations chosen for our Tabu search heuristic,$N_q$ is the number of qubits,  $N_p$ is the number of computers in the quantum network, and \blue{$n$ is the number of non-local binary gates with the chosen home-computer function; note $n \ll N_g$, the number of gates in the circuit}. In our implementation, we pick $\lambda$ to be $20$, beyond which Tabu search offers minimal improvement in solutions for our instances. 

%% file: NetworkTop.tex
\subsection{\bf \nta Step 1: Assignment of Qubits to Computers}
\label{sec:NetworkTop}

Here, we address the first step of \nta --- which \blue{assigns qubits to computers to}
 minimize the
cost of migrations required to cover all the non-local gates.
\blue{In our earlier work~\cite{g2021efficient} where we considered a special case of 
\dqcm problem with homogeneous network and unbounded execution memories,  we used a balanced graph-partitioning to assign qubits to
computers.} However, in the current \dqcm problem, the cost of separating qubits $q_1$ and $q_2$  depends on the specific computers they are assigned to due to the network's 
heterogeneity--- hence, a graph partitioning approach \blue{is inapplicable to} the \dqcm
problem's first step.
\blue{Here, we develop a search-based algorithm---  in particular, based on Tabu search~\cite{GLOVER1986533}--- to assign qubits to computers.}

\para{Tabu Search and Motivation.} Tabu search is a local-search \blue{heuristic} that
starts with an initial solution, and then picks a better solution among
the neighbors of the current solution. To avoid getting stuck in a local minimum, it sometimes also picks a worse solution, especially, if there is no better solution among
the neighbors. The key distinction of Tabu search compared to other local-search algorithms is that it maintains a list of recently-visited solutions and \blue{incurs a penalty each time} one of these solutions is chosen again.
Our motivation for choosing a Tabu-based search heuristic is that our problem closely resembles  the well-studied quadratic-assignment problem for which Tabu search has been shown to perform well~\cite{skorin1990tabu}. \blue{This relationship is clear from our objective function shown under ``Solution' Cost'' below.}

\para{Tabu Search Algorithm for Assignment of Qubits.}
To design a Tabu-search based algorithm for our problem of assignment of qubits to 
computers, we need to define three key aspects of the algorithm: (i) Solution,
(ii) Solution's neighbors, and (iii) Solution's Cost.

\softpara{Solution and Its Neighbors.}
In our context, a solution is a valid home-computer function.
Neighbors of a given solution $\pi$ can be defined as valid solutions $\pi'$ that result from
either: (i) changing the assignment/mapping of a single qubit without violating the storage
constraint, or (ii)  ``swapping'' of two qubits mapped to two different computers in $\pi$.

\softpara{Solution's Cost.}
A solution's cost  can be defined as an
estimate of the cost of migrations needed to cover the non-local gates resulting from the
solution's qubit assignment. 
More formally, the cost of a solution $\pi$, \blue{denoted by $\id{cost}(\pi)$ is} 
$$\id{cost}(\pi) = \sum_{q_1, q_2 \in Q} w(q_1, q_2) \times \text{distance}(\pi(q_1), \pi(q_2))$$
where $w(q_1, q_2)$ is the \emph{number} of migrations needed to cover the binary gates between $q_1$ and $q_2$ if they are assigned to different computers.
We estimate $w(q_1, q_2)$ as described below.

\medskip
\noindent
\emph{Estimating $w(q_1, q_2)$.}
Let ${C}$ is the original circuit that includes two qubits $q_1$ and $q_2$. To estimate 
$w(q_1, q_2)$ in ${C}$,
we consider 
\blue{an induced circuit ${C}'$ that consists only of qubits $q_1$ and $q_2$ and the sequence of gates from ${C}$ that involve $q_1$ and $q_2$.  We can compute the optimal number of migrations} required to cover all the gates in
\blue{the induced circuit ${C}'$} when $q_1$ and $q_2$ are assigned to different computers using the {\em optimal} home-coverage algorithm (called \textsc{MS-HC}) from \cite{g2021efficient}. 
\eat{
small 
\dqcm instance along with a home-computer function as follows. 
We consider two computers with $q_1$ assigned to one of them
and $q_2$ assigned to the other, and consider only those gates from $C$ that
involve $q_1$ and $q_2$ (i.e., binary gates between $q_1$ and $q_2$ in $C$, and
all the unary gates involving them). 
In the above instance, we}
Note that \blue{in ${C}'$} only home-coverage of gates by migrations is possible. 
We use the optimal number of migrations required in \blue{${C}'$} as the estimate
for $w(q_1, q_2)$ in the given circuit ${C}$. 

\softpara{\tabu Algorithm.} \
Based on the above discussion, the 
algorithm (called \tabu) for
the first step of \nta is defined as follows:
\begin{enumerate}
\item $\pi^* = \pi =$ initial random solution
\item $L = [\ ]$ /* a bounded-length list of forbidden solutions */ 
\item Repeat for $\lambda$ iterations:
\begin{enumerate}
\item $\pi = \id{argmin}_{\pi' \in \id{neighbors}(\pi)-L}\ \id{cost}(\pi')$
\item $\pi^* = \pi$ if $\id{cost}(\pi) < \id{cost}(\pi^*)$
\item $L = L \cup \{\pi\}$, removing the oldest element from $L$ if necessary to maintain length bound.
\end{enumerate}
\item Return $\pi^*$
\end{enumerate}

%% file: DynamicPartitioning.tex
\section{\bf General DQC Problem (with Teleportations)}
\label{sec:dynamicPart}


\begin{figure}[t]
\centering
\includegraphics[width=0.45\textwidth]{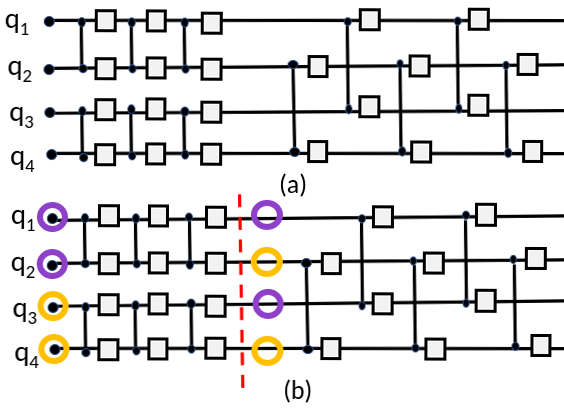}
\eat{\caption{Benefit of using teleportations to execute non-local gates. (a) A quantum circuit instance. (b) The \dqc solution for the instance. Here, the circles signify the computer assigned to the qubit; thus, initially the
set of qubits $\{q_1, q_2\}$ and $\{q_3,q_4\}$ are assigned to the first and second computer respectively,
and then at the red line the qubits assigned to the two computer changes to $\{q_1, q_3\}$ and $\{q_2, q_4\}$.}}
\caption{Circuit illustrating the benefit of including teleportations, see Example~2.}
\vspace*{-1ex}
\label{fig:tele}
\end{figure}
We now consider the general \dqc problem which allows teleportations as well as migrations. We start with illustrating the benefit of teleportations. Then, we design two algorithms for the general \dqc problem; our  algorithms use \nta from the previous section as a subroutine.

\para{Example 2. Benefit of Teleportations.} Consider the circuit instance shown in Fig.~\ref{fig:tele}(a). 
We seek to distribute this circuit across two computers, each with a storage memory of two units. 
If we allow only migrations, then it is easy to observe that the optimal solution assigns the qubits
$\{q_1, q_2\}$ and $\{q_3,q_4\}$ to the two computers respectively and uses five migrations to cover
all the gates (since only 5 of the 11 gates are non-local for the given home-computer mapping). 
Now, consider the solution shown in Fig.~\ref{fig:tele}(b) which uses teleportations too. Here, we essentially change the home-computer mapping at the time instant denoted by the red vertical line,
which makes all the gates local; thus, the total cost is
only 2---for the two teleportations needed to change the home-computer function at the red line. 
\blue{Note that the above example can be scaled to exhibit arbitrarily large the benefit from  using teleportations.}

\begin{figure}[t]
\centering
\includegraphics[width=0.5\textwidth]{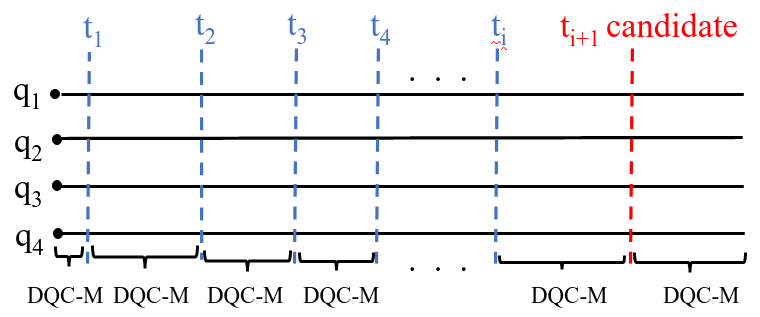}
\caption{\seq algorithm at index $t_{i+1}$. Note that the points of teleportation are determined in sequence from right to left.}
\vspace*{-1ex}
\label{fig:sequenceRun}
\end{figure}
\para{\seq Algorithm.}  Our first algorithm called \seq is a greedy approach, wherein we go over the 
given circuit from left to right, and determine, at each gate, whether or not changing the home-computer function just before it would be beneficial to the overall cost of teleportations and migrations required. 
Recall that we create additional time instants in between gates, and teleportation can only happen at these additional non-gate instants.
Consider a binary gate at time $t+1$. Let us assume that the \seq algorithm has already determined the teleportation points for all the time points before $t-1$. 
To determine whether teleportations should happen at $t$ (i.e., the home-computer function should be changed at $t$), 
we estimate the total cost incurred for the full circuit with optimal teleportations at $t$ \blue{(and none later than $t$)} and compare this estimated cost  with  no teleportations at $t$ \blue{or later}. 
The cost incurred for the full circuit with optimal teleportations at $t$, with $i$ prior instants where teleportations were done, can be estimated in the following way.
\begin{enumerate}[label=(\alph*)]
    \item  Run the \nta algorithm on each of the $i+2$ sub circuits resulting from the $i$ previously chosen instants of teleportation and $t$. 
    \item Compute the total migration cost by adding the migration cost of each sub circuit.
    \item Compute the total teleportation cost by adding the teleportation cost between every pair of consecutive sub circuits.
    \item The cost incurred for the full circuit with optimal teleportations at $t$ is the sum of the total migration cost and total teleportation cost.
\end{enumerate}

The total cost of the whole circuit without teleportation at $t_{i+1}$ can also be similarly estimated (in fact, has already been computed in previous iterations of the algorithm). If the cost with teleportations at $t$ is lower than without the teleportations at $t$, then $t$ is added to the list of time instants where teleportation is to be done.  


\begin{figure}[t]
\centering
\includegraphics[width=0.5\textwidth]{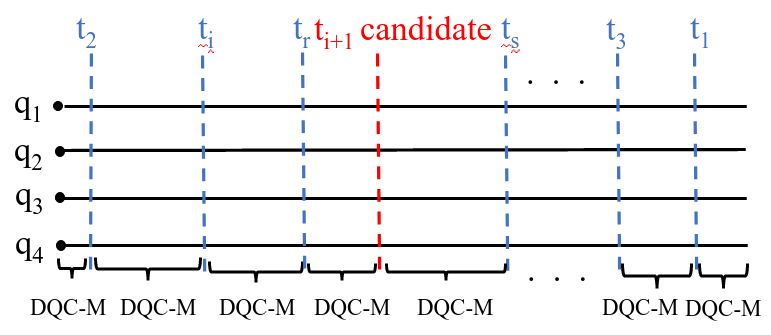}
\caption{\slt algorithm's iteration considering $t_{i+1}$ as the next teleportation-point. Note that the points of teleporation may not be determined in sequence.} 
\vspace*{-1ex}
\label{fig:example3}
\end{figure}

\para{\slt Algorithm.} Our alternate approach to solving the \dqc problem is the \slt algorithm.  \slt is similar to \seq in that \slt also select points of teleportations iterations through a similar cost-estimation methodology. However, rather than going over the circuit from left to right, \slt iteratively picks the best time instant {\em anywhere} in the circuit where teleportation will help the most. 
Consider a stage in the algorithm, where the time instants $t_1, t_2, \ldots, t_i$ have already been determined to be points of teleportations in previous iterations; note that these points need not be ordered left to right. 
Then, in the following iteration, the algorithm  determines the next point
$t_{i+1}$ of teleportation; this 
determination is done by exhaustive search,  by considering all possible points in the circuit and picking the one that yields the best total cost estimate.  The total cost can be estimated in a similar manner as described in the previous \seq algorithm. 
A high-level pseudo-code of \textsc{Split} is as follows.

\begin{itemize}
   \item Assume $i$ points of teleportations $t_1, t_2, \dots , t_i$ have already been chosen; these may not be in left-to-right order. We describe how to select $t_{i+1}$.
   \item For every possible time instant $t$ in the circuit \blue{not in $\{t_1, t_2, \ldots t_i\}$}: 
   \begin{itemize}
       \item Note that choosing $t$ as a point of teleportation results in $i+2$ sub-circuits.
       \item Run \nta on each of these sub-circuits to obtain an initial home-computer function and a migration cost associated with each sub-circuit.
       \item Determine the total estimated cost of picking $t=$ $\sum$ migration costs + $\sum$ teleportation cost of changing the home-computer function. 
   \end{itemize}
   \item Pick the $t$ with minimum cost  as $t_{i+1}$ if and only if it reduces cost from the previous iteration (cost associated with $t_i$). 
   \item Repeat the above steps until no $t$ reduces the cost. 
   \item Run \nta on each sub-circuit to obtain an initial qubit assignment, a set of migrations and a set of teleportations.
\end{itemize}
As in the \seq algorithm, we note that many of the cost components remain unchanged (as the sub-circuits remain unchanged) from one iteration to the next. Hence, these cost components need not be recomputed.

\begin{figure}[t]
\centering
\includegraphics[width=0.5\textwidth]{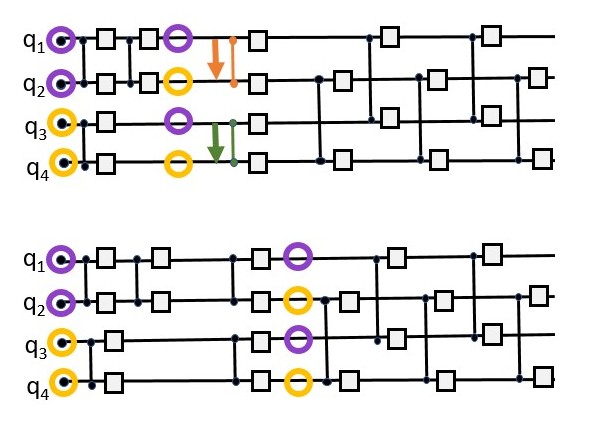}
\caption{An example wherein \slt outperforms \seq.}
\vspace*{-1ex}
\label{fig:splitvseq}
\end{figure}

\begin{figure}[t]
\centering
\includegraphics[width=0.49\textwidth]{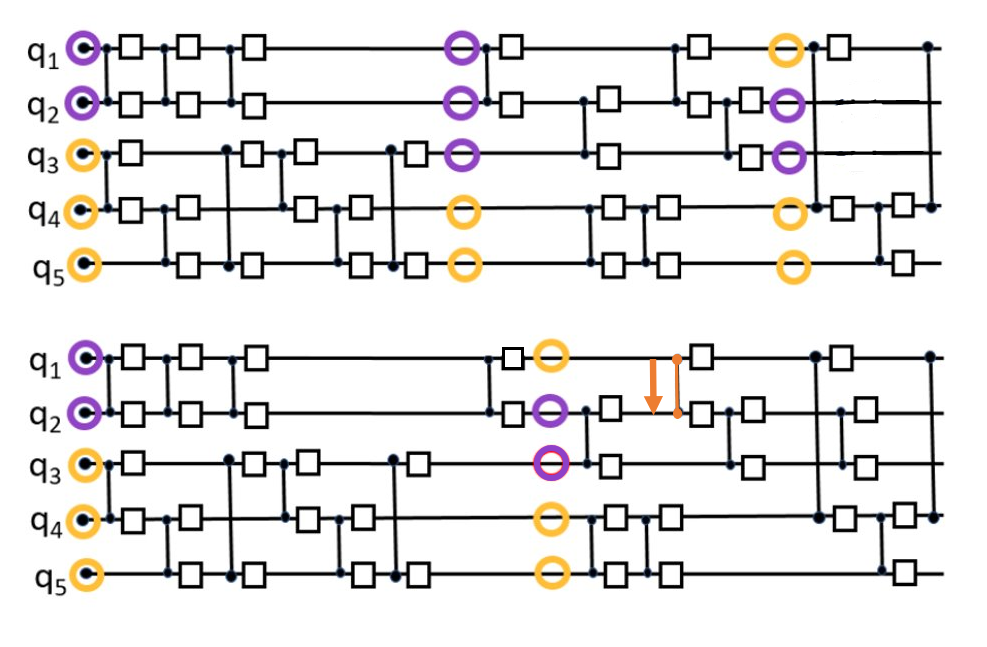}
\caption{An example wherein \seq  outperforms \slt.}
\vspace*{-1ex}
\label{fig:seqvsplit}
\end{figure}

\para{\seq vs.\ \slt Algorithms.}
Since \slt's search methodology is more general, it is expected to outperform \seq; this observation is also confirmed in our empirical results in the next section. However, there are specific instances of the \dqc problem where either may outperform the other. 

Fig.~\ref{fig:splitvseq} shows an instance where \slt outperforms \seq and Fig.~\ref{fig:seqvsplit}) shows an instance where \seq outperforms \slt. In Fig.~\ref{fig:splitvseq}, we observe that \seq tends to pick the earliest point of teleportation that offers an improvement in cost. In contrast, \slt parses the entire circuit and picks the best point of teleportation. This, however, is not always preferable. For example, in Fig.~\ref{fig:seqvsplit}, \slt chooses the best point of teleportation in the first iteration (by arbitrary tie-breaking). This forces \slt to perform one migration and two teleportations. In subsequent iterations, \slt can at best replace the migration with a teleportation, and can never reduce the cost.

\para{Time Complexity of \seq and \slt.} \seq algorithm parses the circuit from left to right, deciding whether or not to teleport at each time instant. Its time complexity is $\mathcal{O}(N_g(T_{DQC\text{-}M}+N_qN_p))$, where $N_q$ is the number of qubits, $N_p$ is the number of computers, $N_g$ is the number of gates in the circuit,
and $T_{DQC\text{-}M}$ is the time taken by the \nta algorithm.
\slt Algorithm repeatedly parses the whole circuit and picks one point of teleportation in each parse. Its time complexity is $\mathcal{O}({N_g}^2(T_{DQC\text{-}M}+N_qN_p))$. In both cases, the term $N_qN_p$ arises from the computation of teleportations.

%% file: Evaluation.tex
\section{\bf Evaluation}
\label{sec:eval}

\begin{figure*}[t]
\begin{subfigure}{\linewidth}
\vspace*{-3.6cm}
    \centering
    \includegraphics[width=0.7\textwidth]{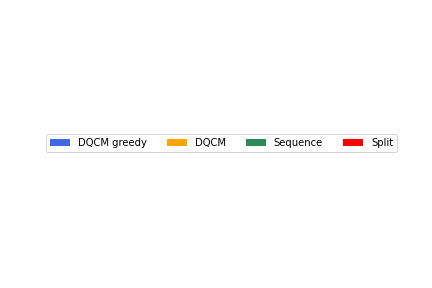}
    \vspace{-3.8cm}
\end{subfigure}
\begin{subfigure}[Varying number of qubits]{0.49\linewidth}
\centering
\includegraphics[width=\textwidth]{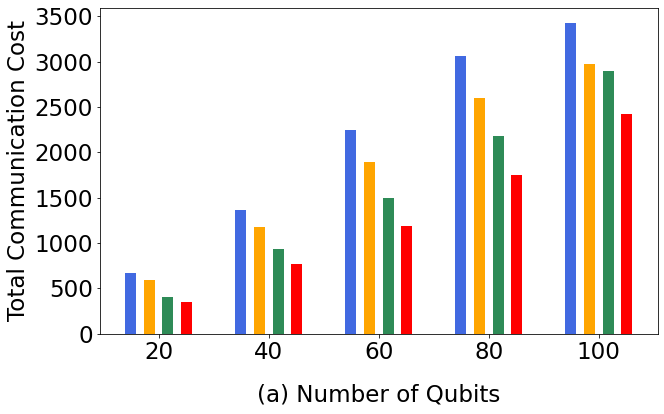}
\captionlistentry{}
\label{fig:VaryQubits}
\end{subfigure}
\begin{subfigure}{0.49\linewidth}
\centering
\includegraphics[width=\textwidth]{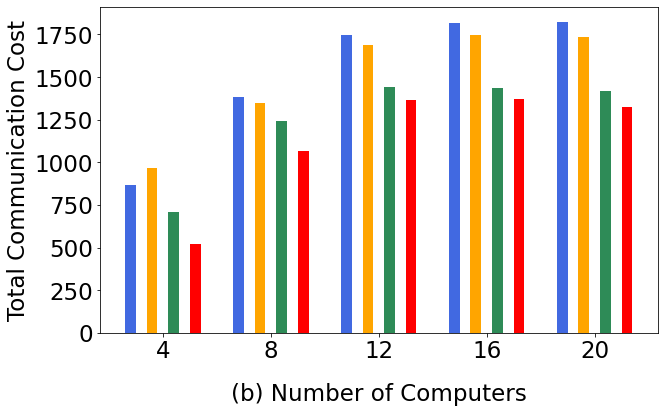}
\captionlistentry{}
\label{fig:VaryPart}
\end{subfigure}
\begin{subfigure}{0.49\linewidth}
\vspace{-0.2cm}
\centering
\includegraphics[width=\textwidth]{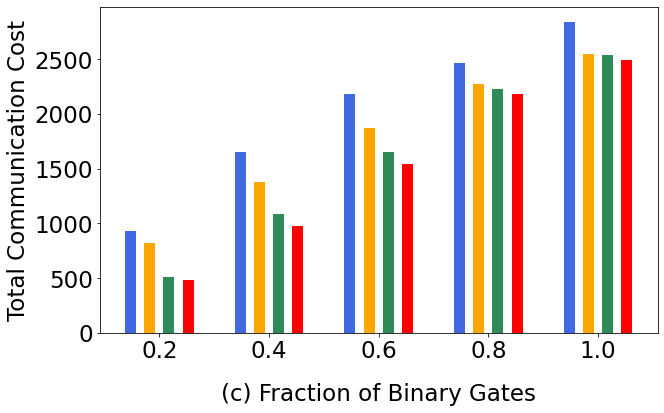}
\captionlistentry{}
\label{fig:VaryFracCZ}
\end{subfigure}
\begin{subfigure}{0.49\linewidth}
\vspace{-0.2cm}
\centering
\includegraphics[width=\textwidth]{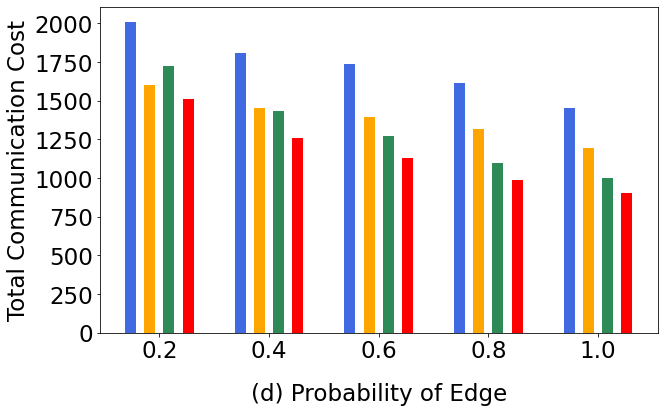}
\captionlistentry{}
\label{fig:VaryProbability}
\end{subfigure}
\begin{subfigure}{0.49\linewidth}
\vspace{-0.2cm}
\centering
\includegraphics[width=\textwidth]{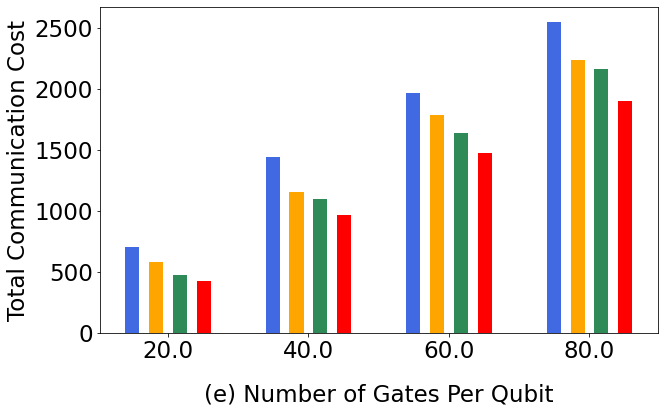}
\captionlistentry{}
\label{fig:VarygateRatio}
\end{subfigure}
\begin{subfigure}{0.49\linewidth}
\vspace{-0.2cm}
\centering
\includegraphics[width=\textwidth]{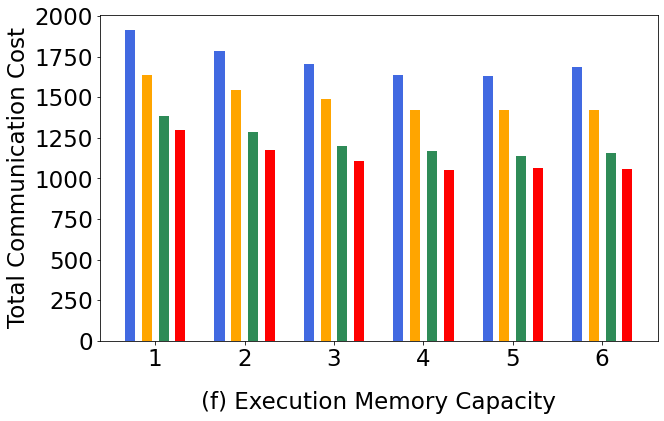}
\captionlistentry{}
\label{fig:VarymemoryConstraints}
\end{subfigure}
\vspace{-0.2cm}
\caption{Total communication cost incurred by different algorithms for varying parameter values.}
\vspace{-0.3cm}
\end{figure*}

Here, we present the evaluation of our algorithms over randomly generated quantum circuits and networks. 
The performance metric used in our evaluations is the the overall communication cost comprising of
the \blue{total} cost of migrations and teleportations as defined before.

\para{Algorithms Compared.} We compare the following four algorithms: (i) \nta from \S\ref{sec:mdqcproblem}, which uses only migrations; (iii) \ntag which is same as \nta, except that it uses a simple greedy algorithm\footnote{An iterative greedy algorithm that selects either a single migration or a pair of migrations that cover the most number of uncovered binary gates without considering execution memory constraints.} instead of \nta's Step 2 to select migrations; 
(iii) \seq from~\S\ref{sec:dynamicPart}  which determines teleportation points by scanning the circuit from left to right; (iv) \slt from~\S\ref{sec:dynamicPart}  which determines teleportations iteratively at arbitrary time instants.


\para{Generating Random Quantum Networks.}  A quantum network is created based on the following parameters.
\begin{itemize}
    \item Number of quantum computers
    \item Probability of a link, between a pair of nodes.
    \item Qubit storage capacity of each computer
    \item Execution memory capacity of each computer
\end{itemize}
To ensure that we generate only connected networks, we use a Python-based library~\cite{networkx} to repeatedly generate Erdős-Rényi graphs with \blue{a given} edge probability until a connected graph is obtained. Erdős-Rényi graphs are generated on $k$ vertices by choosing every edge uniformly at random with probability $p$.  Erdős-Rényi have the useful property that when $p> \frac{(1+\epsilon) \log k}{k}$ for some small $\epsilon>0$, the generated graph is connected with high probability.  In our case, the choices of $k$ and $p$ are sufficient to ensure high probability of connectivity. \eat{We choose $k$, the number of computers, and $p$, the probability of a link, carefully so as to ensure connectivity.} 
We choose the qubit storage capacity of each computer to be $60\%$ to $140\%$ of the average storage requirement (ratio of number of qubits to number of computers). Similarly, we choose the execution memory capacity for each computer to be $30\%$ to $70\%$ of the average storage requirement. 

\para{Generating Random Quantum Circuits.} 
A quantum circuit is created based on the following parameters. 
\begin{itemize}
    \item Number of qubits
    \item Total number of gates (unary and binary) per qubit
    \item Fraction of binary gates, i.e., ratio of binary gates to the total number gates
\end{itemize}
Let 
$f$ be the fraction of binary gates. We generate the gates sequentially, and, at each point, determine whether the gate should be binary (unary) with probability of $f$ ($1-f$). Then, we choose the gate operand(s) randomly.

\para{Evaluation Results.}
We evaluate each of the above four algorithm over generated random networks and circuits as described above. We vary six
parameters in our simulations (the parenthesized values are the corresponding default values): (i) number of computers (10); (ii) probability of an edge (0.5); (iii) number of qubits (50); (iv) number of gates per qubit (50); (v) fraction of binary gates (0.5), (vi) execution memory capacity ($30\%$ to $70\%$ of the average storage requirement).

In each of the experiments, we vary one of the above five parameter values, while fixing the remaining five to their default values. We present the evaluation plots in Figures~\ref{fig:VaryQubits}-\ref{fig:VarygateRatio}. 

Overall,  we make the following observations \blue{on} the relative performances of the algorithms compared.
\begin{itemize}
    \item Using a combination of teleportations and migrations offers a significant reduction in cost 
compared to using only migrations; this is by observing that \seq and \slt algorithms outperform the other two algorithms
in all of our experiments. In particular, in Fig.~\ref{fig:VaryQubits} we see that allowing teleportations reduces the total cost by around $10\%$ on average using \seq and around 
$15\%$ using \slt algorithm.

\item In almost all cases, \slt outperforms \seq; this is as expected, since \slt can be looked upon as a generalization of the \seq algorithm in terms of the candidates considered for teleportation times.

\item
The order of the algorithms from highest to lowest performance is: \slt, \seq, \nta, \ntag; as expected, \ntag performs the worst since it completely disregards the execution-memory constraints initially and resolves the resulting violations in post-processing.
\end{itemize}
We also observed that the \nta algorithm rarely resulted in memory violations, and thus, rarely required any post-processing. 

We also make the following observations regarding how the  
performance of the algorithms varies with varying parameter values. 
In Fig.~\ref{fig:VaryFracCZ}, we observe that with the increase in the ratio of
binary gates, the performance gap between the various algorithms decreases---since with fewer unary gates,  migrations 
don't need to be disentangled much which allows them to 
cover more gates, reducing the advantage of teleportations over migrations. 
In Fig.~\ref{fig:VaryProbability}, we observe that the cost of all algorithms decrease with increasing probability of a network edge, 
due to shorter paths.
In Fig.~\ref{fig:VarygateRatio}, we observe that the cost of all algorithms increase as expected with increase in total number of gates. Finally, in Fig.~\ref{fig:VarymemoryConstraints}, we vary the execution memory capacity of each computer in the quantum network from $1$ to $6$ units, that is, $20\%$ to $120\%$ of average storage requirement (note that the execution memory capacity per computer used in all other plots is $30\%$ to $70\%$ of average storage requirement which amounts to $1$ to $4$ units for each computer).  \blue{As expected, cost decreases with increasing execution memory until there is sufficient memory.}

